\documentclass[12pt]{iopart}

\usepackage{iopams}
\usepackage{amsfonts}
\usepackage{amssymb}
\usepackage{calc}
\usepackage{graphicx}
\usepackage{bm}
\usepackage{color}
\usepackage{citesort}

\def\be{\begin{equation}}
\def\ee{\end{equation}}
\def\bea{\begin{eqnarray}}
\def\eea{\end{eqnarray}}
\def\bse{\begin{subequations}}
\def\ese{\end{subequations}}

\begin{document}

\title{Simulation of Quantum Magnetism in Mixed Spin Systems with Impurity Doped Ion Crystal}

\author{Peter A. Ivanov and Ferdinand Schmidt-Kaler}
\address{Institut f\"ur Physik, Johannes
Gutenberg-Universit\"at Mainz, 55099 Mainz, Germany}

\begin{abstract}
We propose the realization of linear crystals of cold ions which contain different atomic species for investigating quantum phase transitions and frustration effects in spin system beyond the commonly discussed case of $s=\frac{1}{2}$. Mutual spin-spin interactions between ions can be tailored via the Zeeman effect by applying oscillating magnetic fields with strong gradients. Further, collective vibrational modes in the mixed ion crystal can be used to enhance and to vary the strength of spin-spin interactions and even to switch those forces from a ferro- to an antiferromagnetic character. We consider the behavior of the effective spin-spin couplings in an ion crystal of spin-$\frac{1}{2}$ ions doped with high magnetic moment ions with spin $S=3$. We analyze the ground state phase diagram and find regions with different spin orders including ferrimagnetic states. In the most simple non-trivial example we deal with a linear $\{$Ca$^+$, Mn$^+$, Ca$^+\}$ crystal with spins of $\{\frac{1}{2},3,\frac{1}{2}\}$. To show the feasibility with current state-of-the-art experiments, we discuss how quantum phases might be detected using a collective Stern-Gerlach effect of the ion crystal and high resolution spectroscopy. Here, the state-dependent laser-induced fluorescence of the indicator spin-$\frac{1}{2}$ ion, of species $^{40}$Ca$^+$, reveals also the spin state of the simulator spin-$3$ ions, $^{50}$Mn$^+$ as this does not possess suitable levels for optical excitation and detection.
\end{abstract}

 \maketitle

%%%%%%%%%%%%%%%%%%%%%%%%%%%%%%%%%%%%%%%%%%%%%%%%%%%%%%%%%%%%%%%%%%%%%%%
\section{Introduction}\label{introduction}
%%%%%%%%%%%%%%%%%%%%%%%%%%%%%%%%%%%%%%%%%%%%%%%%%%%%%%%%%%%%%%%%%%%%%%%

Current ion trapping technology has led to rapid progress toward the realization of elementary quantum processors \cite{BW,HRB}. The ability to control the motional and  internal states of the trapped ions with high accuracy allows for the experimental implementation of several textbook models such as quantum simulations of a Dirac equation with the Zitterbewegung and the Klein paradox \cite{DiracEq,KParadox}. On the other hand, the internal states of laser-cooled and trapped ions represent effective spins, which can be made to interact with each other for performing magnetic quantum phase simulations. These interactions may be realized by applying magnetic field gradients \cite{CW,CW1,LWRadiation,COspelkaus,2DSS,OS2011} or by laser light fields \cite{PC,Deng,EntanglementTMode,ISLAM2011,LIN2011}. In both cases, spin-$\frac{1}{2}$ ion crystals allow for a detailed investigation of complex quantum phase transitions and magnetic frustrated effects. Preliminary experimental steps in ion crystals have been realized with interacting spins of two $^{25}$Mg$^+$ \cite{QIM1} and three $^{171}$Yb$^+$ ions \cite{QIM2,LIN2011} in a linear configuration. Because of the long-range spin-spin interactions, the larger collections of trapped ions are expected to lead to intriguing, and so far unobserved phenomena, such as the formation of super-solids \cite{XXZ} or exotic quantum phases \cite{BPD}, where the specific advantages of the ion crystal are: almost perfect state preparation and readout with single site addressability, long coherence times, and a full tunability of the spin-spin interactions, even for long ranges beyond next-neighbor couplings.

Going beyond spin-$\frac{1}{2}$ systems and trapping different ion species with spin $S>\frac{1}{2}$ will allow the study of novel aspects of quantum magnetism in a mixed spin chains \cite{QMagnetism,SU}. Such impurity doped systems might model effects which are of interest in solid state physics \cite{ImpuritiesAFM,TransportImpurities}. Our proposal is inspired by the outstanding progress in quantum logic spectroscopy \cite{TR,CHOU2011}, where a single clock ion and a single readout ion are simultaneously confined and coupled through the mutual Coulomb repulsion, such that one can transfer the clock ion electronic state to the readout ion for high fidelity quantum state detection \cite{TM1}. A different, new type of quantum logic readout technique enables us to propose quantum simulation in mixed ion crystals. For the case of neutral interacting atoms, the high magnetic moment of $6\mu_{B}$ of chromium has led to a wealth of novel effects \cite{DipolarEffects,DipolarBEC}, made possible by the tuning of its spin interactions. $^{50}$Mn, with an atomic number which is +1 higher as compared to Cr, shows a similar electronic structure and magnetic moment when singly ionized to Mn$^+$ for being trapped in the ion crystal.

\begin{figure}[tb]
\begin{center}
\includegraphics[width=0.7\linewidth]{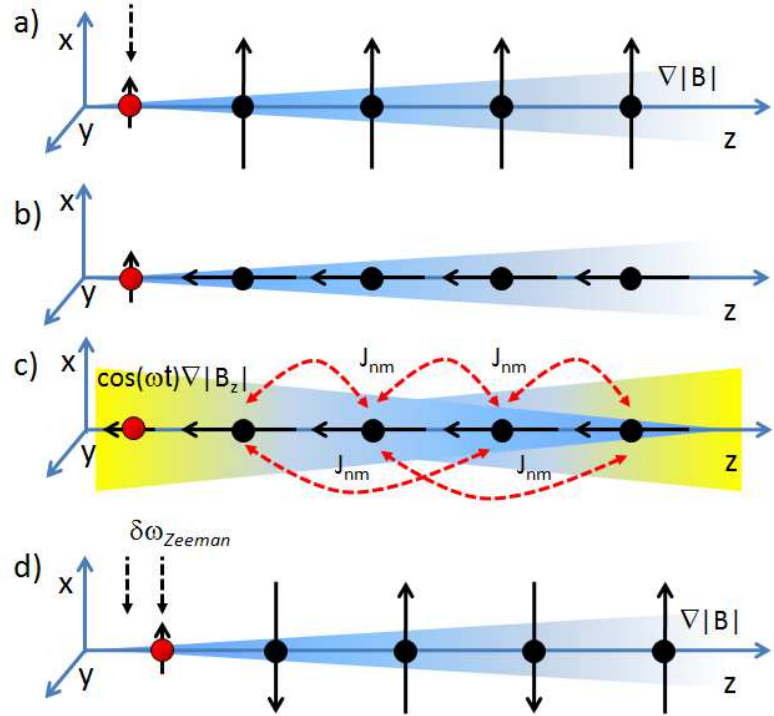}
\end{center}
\caption{Sketch of the proposed experimental sequence which a) initialization in the static magnetic gradient field, followed by a b) RF $\pi/2$ pulse on the Mn$^+$ ions and the c) creation of the spin-spin interactions by using an oscillating magnetic gradient field, and finally concluded by the d) collective spin readout on the Ca$^+$ ion using the position-dependent Stern-Gerlach effect. This leads to a bight or dark state of the Ca$^+$ ion which is imaged on a CCD camera as state-dependent fluorescence.} \label{fig1}
\end{figure}

In this paper, we propose an efficient method for the creation of effective spin-spin interactions in ion crystals of spin-$\frac{1}{2}$ ions doped with different ion species with spin $S=3$. An oscillating magnetic field gradient \cite{COspelkaus} can be used to implement coupling between the spin states of both ion species and the collective motional states of the impurity doped crystal, (see fig.\ref{fig1}). The advantage to use an oscillating magnetic field instead of a laser field is due to avoiding the technical difficulties such as sideband cooling of the many vibrational modes and the necessity to use additional lasers to provide the spin-spin couplings. We show that by proper choice of the frequencies and the direction of the magnetic field gradient, the anisotropic Heisenberg model can be realized with tunable spin-spin couplings. We investigate the particular case of a field gradient applied along the trapping axis such that the spin-spin interactions are described by the transverse Ising model with \emph{single-ion anisotropy}. We consider the ground state phase diagram for a small system consisting of two spin-$\frac{1}{2}$ ions and one spin-$3$ ion placed at the center, which is realizable with the current ion-trap technology. Due to the complex competition between the spin-spin couplings and the single-ion anisotropy we distinguish four regions with different spin orders \cite{LIN2011}. We find that for sufficiently strong antiferromagnetic nearest-neighbor coupling the spin order is \emph{ferrimagnetic} wherein the two different spins are arranged in opposite directions. We show that the ferrimagnetic order can be frustrated due to competing next-nearest-neighbor coupling and the single-ion anisotropy which give rise to a highly entangled ground state.

It has been shown that the atomic gases in optical lattice may be used to realize various condensed-matter models with high spin symmetry \cite{SU,SU2SU6}. The ability to trap ions with large spins at a fixed position \cite{Reordering} and to tune the range and strength of the interactions makes the impurity doped ion crystal an analogue quantum simulator for quantum magnetism and frustration effects in a mixed spin system \cite{FH,Ferrimagnetism}.

The paper is organized as follows: In sec.~\ref{Sec:TM} we describe the theoretical background for the implementation of the effective spin-spin interactions in a $(S,s)=(3,\frac{1}{2})$ mixed spin system by using an oscillating magnetic field gradient. In sec.~\ref{Sec:TIM} we provide analysis of the ground state phase diagram for the transverse Ising Hamiltonian with single-ion anisotropy describing a spin system with two spin-$\frac{1}{2}$ ions and one spin-$3$ ion placed at the center. The method for state preparation and readout of the spin states based on frequency addressing of the auxiliary ion is considered in sec.~\ref{Sec:PSR}. Finally, in sec.~\ref{Sec:conclusion} we give a summary of the results and discuss further and even more complex possibilities of quantum simulation with mixed ion crystals.

%%%%%%%%%%%%%%%%%%%%%%%%%%%%%%%%%%%%%%%%%%%%%%%%%%%%%%%%%%%%%%%%%%%%%%%
\section{Theoretical Model}\label{Sec:TM}
%%%%%%%%%%%%%%%%%%%%%%%%%%%%%%%%%%%%%%%%%%%%%%%%%%%%%%%%%%%%%%%%%%%%%%%

We consider a harmonically confined impurity doped ion crystal with $N-K$ spin-$\frac{1}{2}$ ions with mass $m$ and $K$ spin-$3$ ions with mass $M$. For instance, this is the case of $^{40}$Ca$^+$ ion crystal doped with $^{50}$Mn$^{+}$ ions, which have $^{7}S_{3}$ electronic ground state. If the radial trap frequencies are much larger than the axial trap frequency ($\omega_{x,y}\gg \omega_{z}$), the ions  arrange in a linear configuration along the axial $z$ axis and occupy equilibrium positions \cite{DJames}. The axial trap potential is independent of the mass, so that the equilibrium position of the ions are independent of the composition of the ion crystal. A static magnetic field $B_{0}$ along the trap axis defines the quantization axis. The spin-$\frac{1}{2}$ sublevels $\left\vert \uparrow\right\rangle$ and $\left\vert\downarrow\right\rangle$ are Zeeman split by the applied magnetic field with a resonance frequency $\omega_{0}=(g_{J}\mu_{B}/\hbar)B_{0}$. Here $g_{J}$ denotes the Land\'{e} $g$-factor and $\mu_{B}$ is the Bohr magneton. The spin sublevels of spin-$3$ ions possess seven Zeeman states which we index as $\left\vert m\right\rangle$ with magnetic quantum number $m=-3,-2,\ldots,+3$ and resonance frequency $\tilde{\omega}_{0}$. In the case of ion crystal consisting of $^{40}$Ca$^{+}$ ions with electronic ground state $^{2}S_{1/2}$ doped with $^{50}$Mn$^{+}$ ions the Land\'{e} $g$-factor is $g_{J}\approx 2$ such that the resonance frequencies $\omega_{0}$ and $\tilde{\omega}_{0}$ of both ion species are equal.

\subsection{Magnetic field gradient along the z-direction} \label{Bz}
We assume that the impurity doped ion crystal interact collectively with oscillating magnetic gradient field with frequency $\omega$ applied along the $z$ direction (for simplicity we omit the constant magnetic offset)
\begin{equation}
\vec{B}=\vec{e}_{z}zB_{z}\cos\omega t. \label{Bz}
\end{equation}
The Hamiltonian for $N$ ions interacting with the magnetic field is $\hat{H}=\hat{H}_{0}+\hat{H}_{I}$. Here
\begin{equation}
\hat{H}_{0}=\frac{\hbar\omega_{0}}{2}\sum_{j=1}^{N-K}\sigma_{j}^{z}+\hbar\omega_{0}\sum_{k=1}^{K}S_{k}^{z}+\hbar\sum_{n=1}^{N}\omega_{n,z}\hat{a}_{n,z}^{\dag}\hat{a}_{n,z},\label{H0}
\end{equation}
is the interaction-free Hamiltonian, with $\sigma_{j}^{z}$ being the Pauli matrix for the $j^{th}$ spin-$\frac{1}{2}$ ion and $S_{k}^{z}$ is the spin operator for the $k^{th}$ spin-$3$ ion with $S_{z}|m\rangle=m|m\rangle$. $\hat{a}_{n,z}^{\dag}$ and $\hat{a}_{n,z}$ are, respectively, the creation and annihilation operators of collective phonons along the $z$ axis with frequency $\omega_{n,z}$. The displacement $\hat{z}_{j}$ of the $j^{th}$ ion from its equilibrium position can be expressed in terms of these set of operators as $\hat{z}_{j}=\sum_{n=1}^{N}b_{j,n}^{z}\Delta z_{n}(\hat{a}^{\dag}_{n,z}+\hat{a}_{n,z})$. Here $\Delta z_{n}(a)=\sqrt{\hbar/2a\omega_{n,z}}$  with $a=m,M$ is the spread of the ground state wave function and $b_{j,n}^{z}$ ($n=1,2,\ldots,N$) are the normal mode eigenvectors in the $z$ direction, (see \ref{appendix}) \cite{DJames,SimpCooling}. The interaction between the magnetic dipole moment of the ion species and the magnetic gradient is described by $\hat{H}_{I}=-\hat{\vec{\mu}}.\vec{B}$. The $z$-component of the magnetic-dipole moment for spin-$\frac{1}{2}$ ion is $\hat{\mu}_{z}=(\gamma/2)\sigma_{z}$ and, respectively, for spin-$3$ $\hat{\mu}_{z}=\gamma S_{z}$ with $\gamma=\mu_{B}g_{L}$. We may transform the Hamiltonian in the interaction picture with respect to $\hat{H}_{0}$ to obtain
%========================================================================
\begin{figure}[tb]
\begin{center}
\includegraphics[width=1.0\linewidth]{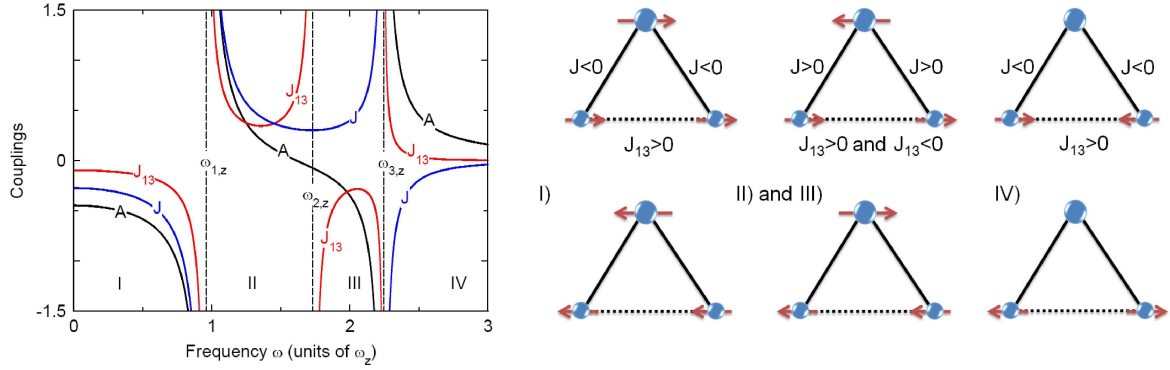}
\end{center}
\caption{left). The nearest-neighbor $J_{12}=J_{23}=J$, the next-nearest-neighbor $J_{13}$ spin-spin couplings and the single-ion anisotropy $A$ as a function of $\omega$ for ion crystal consisting of two spin-$\frac{1}{2}$ ions and one spin-$3$ ion placed at the center, (see Eq. (\ref{SC})). The couplings $J$, $J_{13}$ and $A$ are normalized to $\varepsilon=(\Delta z \partial_{z}\omega_{0})^{2}/(2\omega_{z})$, which quantifies the change of the spin resonance frequency $\omega_{0}$ due to the shift of the equilibrium position of the ion with trap frequency $\omega_{z}$ by an amount equal to the spread of the ground state wavefunction, $\Delta z=\sqrt{\hbar/2m\omega_{z}}$. right). Quantum phases differ in the regions (I) to (IV).} \label{fig2}
\end{figure}
%========================================================================

\begin{eqnarray}
\hat{H}_{I}^{z}=-\hbar\sum_{n=1}^{N}\left(\sum_{j=1}^{N-K}\frac{\Omega_{j,n}^{z}}{2}\sigma_{j}^{z}+\sum_{k=1}^{K}\Omega_{k,n}^{z}S_{k}^{z}\right)\left(\hat{a}_{n,z}^{\dag}e^{i(\omega+\omega_{n,z})t}+\hat{a}_{n,z}e^{-i(\omega+\omega_{n,z})t}
\right)\nonumber \\
-\hbar\sum_{n=1}^{N}\left(\sum_{j=1}^{N-K}\frac{\Omega_{j,n}^{z}}{2}\sigma_{j}^{z}+\sum_{k=1}^{K}\Omega_{k,n}^{z}S_{k}^{z}\right)\left(\hat{a}_{n,z}^{\dag}e^{-i(\omega-\omega_{n,z})t}+\hat{a}_{n,z}e^{i(\omega-\omega_{n,z})t} \right).\label{H}
\end{eqnarray}
The function $\Omega_{j,n}^{z}=b_{j,n}^{z}\Delta z_{n}B_{z}\gamma/2\hbar$ is the Rabi frequency of the $j^{th}$ ion, which quantifies the coupling to the $n^{th}$ vibrational mode. Hence, the oscillating magnetic gradient field mediates a coupling between the internal states of the ions and the external (motional) states of the ion crystal. Indeed, the two terms in the Hamiltonian (\ref{H}) describe a time-varying spin-dependent displacement with frequencies ($\omega+\omega_{n,z}$) and ($\omega-\omega_{n,z}$). If the frequency $\omega$ is not resonant to any vibrational mode and the condition $|\omega_{n,z}-\omega|\gg \Omega_{j,n}^{z}$ is satisfied for any $n$ then we can perform time-averaging of the rapidly oscillating terms in (\ref{H}) \cite{DJames1}. Hence, we arrive at the following time-averaged effective Hamiltonian
\begin{eqnarray}
\hat{H}_{eff}^{z}&=&\hbar\sum_{j,j'=1\atop j>j^{'}}^{N-K}J_{j,j^{'}}^{(1,z)}\sigma_{j}^{z}\sigma_{j'}^{z}+\hbar\sum_{k,k'=1\atop k>k^{'}}^{K}J_{k,k'}^{(2,z)}S_{k}^{z}S_{k^{'}}^{z}\nonumber \\
&&+\hbar\sum_{j,k=1}^{N}J_{j,k}^{(3,z)}\sigma_{j}^{z}S_{k}^{z}+\hbar\sum_{k=1}^{K}A_{k}^{z}(S_{k}^{z})^{2}.\label{Heff}
\end{eqnarray}
Therefore, the off-resonant oscillating magnetic gradient creates effective spin-spin interaction between the identical \cite{ISLAM2011,LIN2011} and different ion species in the crystal. The first two terms in (\ref{Heff}) quantify the spin-spin coupling between the spin-$\frac{1}{2}$ ions and the spin-$3$ ions. The third term in (\ref{Heff}) describes the spin-spin coupling between the different ion species. Surprisingly, the adiabatic elimination of the vibrational modes for a ion crystal with $s>\frac{1}{2}$ ions gives rise to single-ion anisotropy term $A_{j}^{z}$ which quantifies the non-linear Zeeman shift of the spin-$3$ magnetic sublevels. We note that the non-linear Zeeman shift for spin-$\frac{1}{2}$ ions is equal for the both magnetic sublevels, thereby it is proportional to the unit matrix.  The couplings in (\ref{Heff}) are given by
\begin{eqnarray}
J_{j,j'}^{(1,z)}=\frac{B_{z}^{2}\gamma^{2}}{8\hbar m}\sum_{n=1}^{N}\frac{b_{j,n}^{z}b_{j^{'},n}^{z}}{\omega^{2}-\omega_{n,z}^{2}},\, J_{k,k'}^{(2,z)}=\frac{B_{z}^{2}\gamma^{2}}{2\hbar M}\sum_{n=1}^{N}\frac{b_{k,n}^{z}b_{k^{'},n}^{z}}{\omega^{2}-\omega_{n,z}^{2}},\nonumber \\
J_{j,k}^{(3,z)}=\frac{B_{z}^{2}\gamma^{2}}{4\hbar \sqrt{m M}}\sum_{n=1}^{N}\frac{b_{j,n}^{z}b_{k,n}^{z}}{\omega^{2}-\omega_{n,z}^{2}},\, A_{k}^{z}=\frac{B_{z}^{2}\gamma^{2}}{4\hbar M}\sum_{n=1}^{N}\frac{(b_{k,n}^{z})^{2}}{\omega^{2}-\omega_{n,z}^{2}}.\label{SC}
\end{eqnarray}
The main advantage of using an oscillating magnetic field gradient instead of constant is that we may engineer a variety of interactions between the ions. Fig.~\ref{fig2} shows the spin-spin couplings and the single-ion anisotropy (\ref{SC}) for a chain of two spin-$\frac{1}{2}$ ions and one spin-$3$ ion place at the center versus the frequency $\omega$. In contrast with the constant magnetic field gradient applied along the trapping axis $z$ wherein the spin-couplings can be only ferromagnetic, now as $\omega$ is varied the magnitude and the sign of the couplings are changed which allows the creation of ferromagnetic, antiferromagnetic or frustrated interaction between the ions. The ground state of Hamiltonian (\ref{Heff}) highly depends from the sign of the single-ion anisotropy terms $A_{k}^{z}$. Indeed, for sufficiently large positive single-ion anisotropy $(A_{k}^{z}\gg 0)$, the spin-$3$ ions have magnetic quantum number $m=0$ for the ground state, while in the opposite limit $(|A_{k}^{z}|\gg 0)$ the spin-$3$ ground state projection is $m=\pm 3$.

\subsection{Magnetic field along xyz-direction}\label{Bxyz}
%========================================================================
\begin{figure}[tb]
\begin{center}
\includegraphics[angle=0,width=0.9\linewidth]{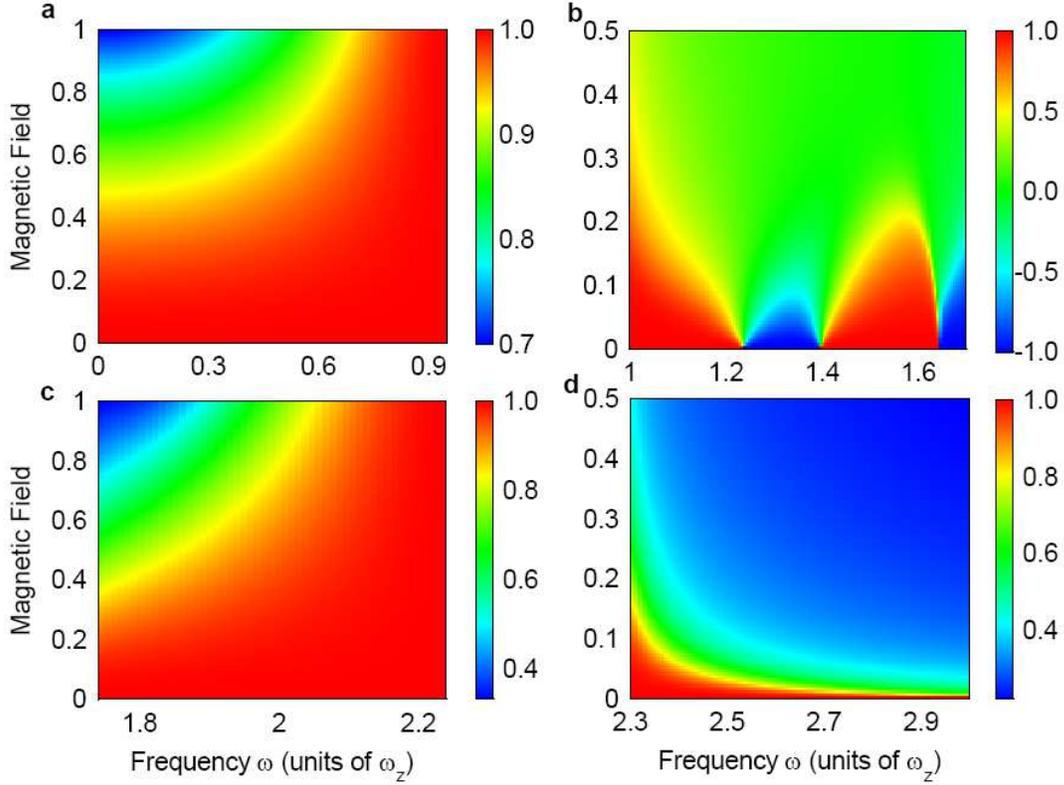}
\end{center}
\caption{The ground state phase diagram calculated by an exact diagonalization of Hamiltonian (\ref{TI}). a) The ferromagnetic population $P_{fm,3}$ as a function of the normalized transverse magnetic field $B_{x}^{0}/\varepsilon$ and the frequency $\omega$. In region I) the spin-spin couplings are ferromagnetic $J,J_{13}<0$ and the single-ion anisotropy is $A<0$. b) The population difference $P_{f,1}-P_{f,2}+P_{f,3}-
P_{a,3}$ as a function of the normalized transverse magnetic field $B_{x}^{0}/\varepsilon$ and the frequency $\omega$. In region II) the spin-spin couplings are antiferromagnetic $J,J_{13}>0$ and the single-ion anisotropy is $A>0$. By increasing $\omega$ as $B_{x}^{0}\rightarrow 0$ the system undergoes transition $\left\vert\psi_{f,1}\right\rangle \rightarrow \left\vert\psi_{f,2}\right\rangle \rightarrow \left\vert\psi_{f,3}\right\rangle \rightarrow \left\vert\psi_{a,3}\right\rangle$. c) The ferrimagnetic population $P_{f,3}$ as a function of the normalized transverse magnetic field $B_{x}^{0}/\varepsilon$ and the frequency $\omega$. In region III) the spin-spin couplings are, respectively, $J>0$, $J_{13}<0$ and the single-ion anisotropy is $A<0$. d) The antiferromagnetic population $P_{a,0}$ as a function of the normalized transverse magnetic field $B_{x}^{0}/\varepsilon$ and the frequency $\omega$. In region IV) the spin-spin couplings are $J<0$, $J_{13}>0$, respectively, and the single-ion anisotropy is $A>0$. } \label{fig3}
\end{figure}
%========================================================================
Consider the magnetic gradient applied along the $xyz$ direction
\begin{equation}
\vec{B}(x,y,z)=\vec{e}_{z}zB_{z}\cos \omega t + \vec{e}_{x}xB_{x}f(t)-\vec{e}_{y}yB_{y}f(t), \label{Bxyz}
\end{equation}
with $f(t)=(\cos \omega_{b} t +\cos \omega_{r} t)$. Such a field can be created in a micro structured planar ion trap, which contains a central wire loop \cite{2DSS}. The oscillating field in $x$-$y$ direction provides additional coupling $\Omega_{j,n}^{q}=b_{j,n}^{q}\Delta q_{n}B_{q}\gamma/2\hbar$ ($q=x,y$) between the internal and motional degree of freedom of the ion crystal. We assume that the frequencies $\omega_{b}-\omega_{0}=\delta$ and $\omega_{r}-\omega_{0}=-\delta$ are tuned to the blue- and red- sideband transition with detuning $\pm \delta$. Then the $x$-$y$ gradients induce a spin-dependent displacement with frequencies $(\delta+\omega_{n,q})$ and $(\delta-\omega_{n,q})$ similar to Eq. (\ref{H}), where $\omega_{n,q}$ are the vibrational frequencies, respectively, in the $x$ and $y$ directions, (see \ref{appendix}). After applying an optical rotating-wave approximation (neglecting the terms $\omega_{0}+\omega_{b,r}$) and assuming that $|\omega_{n,q}-\delta|\gg \Omega_{j,n}^{q}$ is fulfilled for any vibrational mode in the $x$-$y$-direction, the time-averaged effective Hamiltonian is given by
\begin{eqnarray}
\hat{H}_{eff}&=&\hbar\sum_{q=x,y,z}\{\sum_{j,j^{'}=1\atop j>j^{\prime}}^{N-K}J_{j,j^{'}}^{(1,q)}\sigma_{j}^{q}\sigma_{j'}^{q}+\sum_{k,k^{\prime}=1\atop k>k^{\prime} }^{K}J_{k,k'}^{(2,q)}S_{k}^{q}S_{k^{'}}^{q}\nonumber \\
&&+\sum_{j,k=1}^{N}J_{j,k}^{(3,q)}\sigma_{j}^{q}S_{k}^{q}+\sum_{k=1}^{K}A_{k}^{q}(S_{k}^{q})^{2}\}.\label{Hxyz}
\end{eqnarray}
The spin-spin couplings in $x$-$y$ direction are identical in form to (\ref{SC}) by replacing $\omega\rightarrow \delta$ and $z\rightarrow x, y$. Hence, the magnetic field (\ref{Bxyz}) creates an anisotropic Heisenberg ($XYZ$) interaction between the effective spins.

\section{Transverse Ising Model}\label{Sec:TIM}
The quantum transverse Ising Hamiltonian is given by
\begin{equation}
H_{TI}=H_{eff}^{z}-\hbar B_{x}^{0}\{\sum_{j=1}^{N-K}\frac{\sigma_{j}^{x}}{2}+\sum_{k=1}^{K}S_{k}^{x}\}.\label{TI}
\end{equation}
The last term in (\ref{TI}) can be simulated by driving transitions between the ion spin states employing radio frequency field $\vec{B}_{0}=\vec{e}_{x}B_{0}\cos\tilde{\omega} t$. Assuming the resonance condition is fulfilled, i.e., $\tilde{\omega}=\omega_{0}$ we obtain the effective transverse field $B_{x}^{0}=\gamma B_{0}/2\hbar$.

The simplest non-trivial case is to consider ion chain with two spin-$\frac{1}{2}$ ions and one spin-$3$ ion placed at the center. Such ordering of ions is consistent with the natural behavior, as observed in Ref.~\cite{Reordering}. When applying the oscillating gradient field, the resulting spin-spin couplings are shown in fig.~\ref{fig2} as a function of the drive frequency $\omega$. We may distinguish four different regions wherein the spin-spin interactions are ferromagnetic, antiferromagnetic, or frustrated. The presence of the single-ion anisotropy $A$ in (\ref{TI}) changes substantially the ground state phase diagram compared to the case of spin-$\frac{1}{2}$ ion chain. In contrast to the spin-$\frac{1}{2}$ string, the ground state of the mixed $(S,s)=(3,\frac{1}{2})$ spin system can be frustrated due to the complex competition between the spin-spin couplings $J$ and $J_{13}$ and the single-ion anisotropy $A$.

In region I), see fig.~2, all interactions are ferromagnetic $(J,J_{13}<0)$ and the resulting ground state of the Hamiltonian (\ref{TI}) as $B_{x}^{0}\rightarrow 0$ is a coherent superposition of two ferromagnetic states $\left\vert\uparrow\uparrow\right\rangle\left\vert 3 \right\rangle$ and $\left\vert\downarrow \downarrow \right\rangle \left\vert-3\right\rangle$. Because the single-ion anisotropy is negative $(A<0)$, the ground state energy is minimized for spin-$3$ state with magnetic quantum number $m=\pm 3$. The ferromagnetic population $P_{fm,3}=P_{\uparrow\uparrow 3}+P_{\downarrow \downarrow -3}$ as a function of the effective magnetic field and the frequency $\omega$ is shown in fig.~\ref{fig3}a. In the region II) the spin interactions are antiferromagnetic $(J,J_{13}>0)$. This is the case of ferrimagnetism in which spins of two types interact by nearest-neighbor antiferromagnetic coupling, (see fig.~\ref{fig3}b). However, the ferrimagnetic interaction is frustrated due to the competing next-nearest-neighbor antiferromagnetic coupling $J_{13}>0$ which disturbs the ferrimagnetic order and tends to align the two spins-$\frac{1}{2}$ in a antiferromagnetic state. Additionally, the ferrimagnetic interaction is also frustrated due to the strong positive single-ion anisotropy $A>0$ which attempts to project the spin-$3$ state $\left\vert m \right\rangle$ in a quantum number $m=0$. In the beginning of region II) $A>0$ is high but the ferrimagnetic configuration is still energetically favorable such that the ground state of Hamiltonian (\ref{TI}) as $B_{x}^{0}\rightarrow 0$ is a superposition of two ferrimagnetic states $\left\vert\psi_{f,1}\right\rangle=(\left\vert\uparrow\uparrow\right\rangle\left\vert -1 \right\rangle + \left\vert\downarrow \downarrow \right\rangle \left\vert 1\right\rangle)/\sqrt{2}$, wherein the spins $S=3$ and $s=\frac{1}{2}$ are aligned anti-parallel with each other. By increasing $\omega$ the single-ion anisotropy $A$ decreases which allows the quantum number $m$ to increase and the resulting ground state is $\left\vert\psi_{2,f}\right\rangle =(\left\vert\uparrow\uparrow\right\rangle\left\vert -2 \right\rangle +\left\vert\downarrow \downarrow \right\rangle \left\vert 2\right\rangle)/\sqrt{2}$. For $A/J=2/3$ which occurs at $\omega\approx1.24\omega_{z}$ the ground state for $B_{x}^{0}=0$ is an entangled superposition of four ferrimagnetic states $\left\vert \uparrow \uparrow\right\rangle \left\vert -1\right\rangle$, $\left\vert \downarrow \downarrow\right\rangle \left\vert 1\right\rangle$, $\left\vert \uparrow \uparrow\right\rangle \left\vert -2\right\rangle$, and $\left\vert \downarrow \downarrow\right\rangle \left\vert 2\right\rangle$. By decreasing $A$ the ferrimagnetic state $\left\vert\psi_{2,f}\right\rangle$ undergoes a transition to $\left\vert\psi_{3,f}\right\rangle=(\left\vert\uparrow\uparrow\right\rangle\left\vert -3 \right\rangle +\left\vert\downarrow \downarrow \right\rangle \left\vert 3\right\rangle)/\sqrt{2}$. At the transition point $\omega=1.4\omega_{z}$ and $A/J=0.4$ the resulting ground state for $B_{x}^{0}=0$ is an entangled superposition of four ferrimagnetic states, $\left\vert \uparrow \uparrow\right\rangle \left\vert -2\right\rangle$, $\left\vert \downarrow \downarrow\right\rangle \left\vert 2\right\rangle$, $\left\vert \uparrow \uparrow\right\rangle \left\vert -3\right\rangle$, and $\left\vert \downarrow \downarrow\right\rangle \left\vert 3\right\rangle$. For sufficiently high positive $J_{13}$ the ferrimagnetic arrangement is not favorable and the two spin-$\frac{1}{2}$ ions are arranged in an antiferromagnetic state. Hence, the ferrimagnetic configuration is broken and the ground state is $\left\vert\psi_{a,3}\right\rangle =(\left\vert\uparrow\downarrow\right\rangle +\left\vert\downarrow \uparrow \right\rangle)(\left\vert 3 \right\rangle + \left\vert -3 \right\rangle)/2$. At the transition point $\omega\approx1.65\omega_{z}$, the ground state of the Hamiltonian (\ref{TI}) is an entangled superposition of six states: two ferrimagnetic states $\left\vert \uparrow \uparrow\right\rangle \left\vert -3 \right\rangle$ and $\left\vert \downarrow \downarrow\right\rangle \left\vert 3 \right\rangle$, and four antiferromagnetic states $\left\vert \uparrow \downarrow\right\rangle \left\vert 3 \right\rangle$, $\left\vert \uparrow \downarrow\right\rangle \left\vert -3 \right\rangle$, $\left\vert \downarrow \uparrow\right\rangle \left\vert 3 \right\rangle$, and $\left\vert \downarrow \uparrow\right\rangle \left\vert -3 \right\rangle$. In region III) the nearest-neighbor spin coupling is antiferromagnetic $(J>0)$ such that the ferrimagnetic configuration is favorable. In contrast with the region II), no frustration exist because the next-nearest-neighbor coupling is ferromagnetic $J_{13}<0$ and $A<0$. In the entire region the ground state is a ferrimagnetic $\left\vert\psi_{f,3}\right\rangle$, which minimizes all interactions. In fig.~\ref{fig3}c the ferrimagnetic population $P_{f,3}=P_{\uparrow\uparrow -3}+P_{\downarrow \downarrow 3}$ versus the effective magnetic field and the frequency $\omega$ is shown . In region IV) the single-ion anisotropy is positive ($A>0$) which causes the spin-$3$ being projected into quantum number $m=0$. Additionally, the next-nearest-neighbor coupling is antiferromagnetic $J_{13}>0$ and the resulting ground state as $B_{x}^{0}\rightarrow 0$ is antiferromagnetic $\left\vert\psi_{a,0}\right\rangle =(\left\vert\uparrow\downarrow\right\rangle\ +\left\vert\downarrow \uparrow \right\rangle)\left\vert 0 \right\rangle/\sqrt{2}$. The antiferromagnetic population $P_{a,0}=P_{\uparrow\downarrow 0}+P_{\downarrow \uparrow 0}$ is shown in fig.~\ref{fig3}d.

\begin{figure}[tb]
\begin{center}
\includegraphics[angle=0,width=0.9\linewidth]{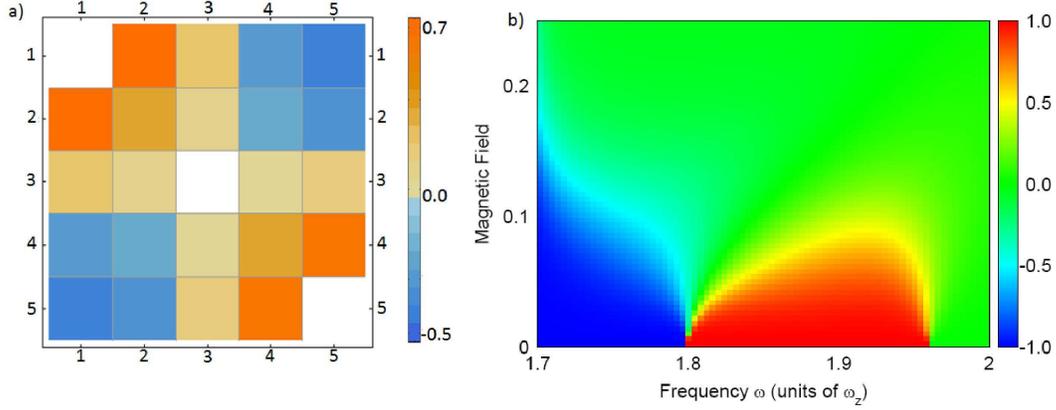}
\end{center}
\caption{a). The spin-spin couplings and single-ion anisotropy in alternating mixed-spin chain $\{\frac{1}{2},3,\frac{1}{2},3,\frac{1}{2}\}$ consisting of three spin-$\frac{1}{2}$ ions and two spin-$3$ ions. All couplings are normalized to $\varepsilon$. The frequency of the oscillating magnetic field is set to $\omega=1.8\omega_{z}$. The nearest-neighbor couplings are antiferromagnetic. Similar antiferromagnetic behavior is observed in larger alternating odd number ion crystals. b) Because of the antiferromagnetic nearest-neighbor interactions, the different spins are aligned anti-parallel with each other and the resulting ground state as $B_{x}^{0}/\varepsilon \rightarrow 0$ is ferrimagnetic $\left\vert\psi_{f,2}\right\rangle=(\left\vert \uparrow \uparrow \right\rangle \left\vert -2,-2 \right\rangle+\left\vert \downarrow\downarrow \right\rangle \left\vert 2,2 \right\rangle)/\sqrt{2}$ or $\left\vert\psi_{f,3}\right\rangle=(\left\vert \uparrow \uparrow \right\rangle \left\vert -3,-3 \right\rangle+\left\vert \downarrow\downarrow \right\rangle \left\vert 3,3 \right\rangle)/\sqrt{2}$. The figure shows the population difference $P_{f,3}-P_{f,2}$ as a function of $\omega$ and $B_{x}^{0}/\varepsilon$.} \label{fig4}
\end{figure}

\subsection{Towards larger and more complex mixed crystal}
By increasing the number of ions the spin-spin interactions become more complex and consequently the new ground states exhibit a variety of spin orders. In alternating mixed-spin chain we can find regions in which the nearest-neighbor interactions are antiferromagnetic with ferrimagnetic ground states, (see fig. \ref{fig4}). Such antiferromagnetic mixed-spin chains have been proposed as a model for describing certain molecular-based magnets of experimental interest \cite{Willem}. Moreover, one of the most studied topics in alternating mixed-spin chains is the appearance of a plateaux in the magnetization curve \cite{Ferrimagnetism,SAKAI,Yamamoto,OSHIKAWA}. It has been shown that the formation of these plateaux depend highly on the competing interactions and single-ion anisotropy. Thus, the proposed ion trap based simulator of quantum magnetism in mixed-spin system will allow for detailed studies of such phenomena.

%%%%%%%%%%%%%%%%%%%%%%%%%%%%%%%%%%%%%%%%%%%%%%%%%%%%%%%%%%%%%%%%%%%%%%%%
\section{Preparation and Spin Readout}\label{Sec:PSR}
%%%%%%%%%%%%%%%%%%%%%%%%%%%%%%%%%%%%%%%%%%%%%%%%%%%%%%%%%%%%%%%%%%%%%%%%

\begin{figure}[tb]
\begin{center}
\includegraphics[width=0.7\linewidth]{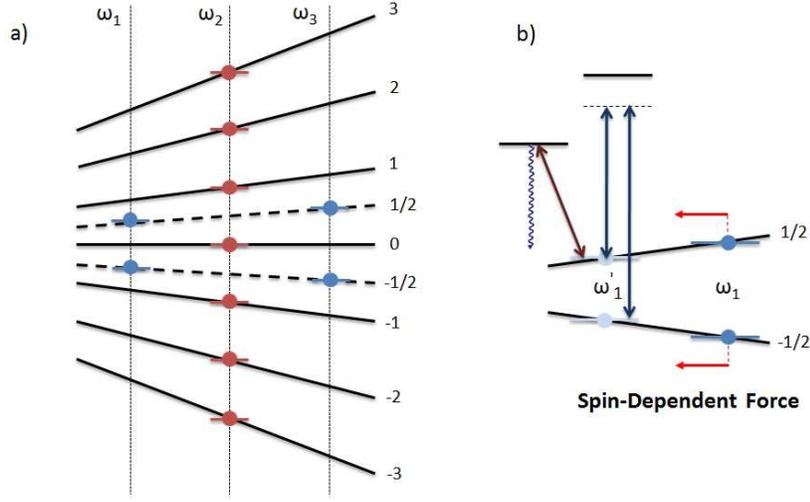}
\end{center}
\caption{a) In the presence of static magnetic field gradient the Zeeman shift of the spin states is different for each ion. b) The ion equilibrium position is shifted due to the spin-dependent force acting on the magnetic dipole moment. The direction and the magnitude of the displacement depends on the spin states of the other ions. Thus, the spin resonance frequency will change from $\omega_{1}$ to $\omega_{1}^{'}$. By scanning the frequency of the applied laser light the spin transition comes into resonance and consequently the ion scatter light.} \label{fig5}
\end{figure}

In order to prepare the initial state of the Hamiltonian
\begin{equation}
\hat{H}_{TI}^{0}=-\hbar B_{x}^{0}\{\frac{\sigma_{1}^{x}}{2}+S_{2}^{x}+\frac{\sigma_{3}^{x}}{2}\} \label{H0TI}
\end{equation}
and to readout the final state after the adiabatic ramping of the spin couplings we can apply a static magnetic field gradient $\vec{B}=\vec{B}_{0}+bz\vec{e}_{z}$ along the trapping axis $z$ which shifts the resonance frequencies of transitions between Zeeman states, (see fig.~\ref{fig5}a and the general scheme in fig.~\ref{fig1}). Due to the spatial variation of the magnetic field, the spin states exhibit site-specific resonance frequency, $\omega_{0}^{'}=\omega_{0}+\gamma b z/\hbar$. Due to the applied gradient field, the equilibrium position of each of the ions is shifted depending of the spin states of the other ions in the linear crystal. Indeed, the total force which acts on the magnetic dipole moment of spin-$\frac{1}{2}$ ion along the trapping axis $z$ is $F_{1}=F_{z}^{(2)}+F_{z}^{(3)}$, where $F_{z}^{(2)}=\hbar\partial_{z}\omega\langle S_{z}\rangle$ and $F_{z}^{(3)}=(\hbar/2)\partial_{z}\omega\langle \sigma_{z}\rangle$ are forces associated with the spins $\hbar S_{z}$ and $(\hbar/2)\sigma_{z}$ in a magnetic field $\vec{B}$ and $\langle \rangle$ indicates the expectation value \cite{CW}. The spin-dependent force shifts the equilibrium position by an amount $d_{z}=(F_{z}^{(2)}+F_{z}^{(3)})/(m\omega_{z}^{2})$. Consequently, the spin resonance frequency is changed from $\omega_{1}$ to $\omega_{1}^{'}$, with $\omega_{1}^{'}=\omega_{1}+\gamma b d_{z}/\hbar$, (see fig.~\ref{fig5}b). The frequency separation $\delta \omega_{Zeeman}=|\omega_{1}^{'}-\omega_{1}|$ between the two spin resonances of spin-$\frac{1}{2}$ ion is given by
\begin{equation}
\delta \omega_{Zeeman} = \frac{\mu_{B}g_{J}b}{\hbar}|d_{z}|.\label{SR}
\end{equation}
Assuming a high but realistic magnetic gradient of $b=20$ T/m \cite{OS2011} and an axial trap frequency of $\omega_{z}=2\pi\times 100$~kHz, the shift from the equilibrium position of a spin-$\frac{1}{2}$ ion corresponding to the state $\left\vert m=3 \right \rangle_{2} \left\vert \uparrow \right \rangle_{3}$ is approximately $d_{z}\approx50$~nm which gives rise to frequency separation (\ref{SR}) of about $\delta \omega_{Zeeman} \approx 174$ kHz. Note that much smaller frequency shifts are easily resolved on a narrow Raman or quadrupole transition, such that in a final step the observation of laser-induced fluorescence allows one to determine the spin states.
%========================================================================
%========================================================================
The experimental sequence is started by preparing the two spin-$\frac{1}{2}$ ions in a state $\left\vert \downarrow \downarrow \right\rangle \left\vert \Psi_{S} \right\rangle$, where $\left\vert \Psi_{S} \right\rangle$ is a still unknown state of the spin-$3$ ion. Additionally, we may trap an auxiliary ion initially prepared by optical pumping in the state $\left\vert \downarrow \right\rangle_{a}$  which can be used to readout the spin states of the remaining three ions.  In the second step we switch on the static magnetic gradient, which creates the spin-dependent force. Next, we expose the auxiliary ion to laser light with frequency $\omega_{1}^{'}$, which drives the transition $\left\vert \downarrow \right\rangle_{a} \leftrightarrow \left\vert \uparrow \right\rangle_{a}$ only if the state of spin-$3$ ion is $\left\vert m=-3 \right\rangle$. If the auxiliary ion does not scatter light we have to discard the measurement and restart. If we observe fluorescence then the state of the spin-$3$ ion is measured in $\left\vert -3 \right\rangle$ and results, after performing a $\pi/2$ rotation along the $y$ axis, is the desired ground state $\left\vert \Psi_{0}\right\rangle$ of Hamiltonian (\ref{H0TI})
\begin{equation}
\left\vert \Psi_{0}\right\rangle = e^{-i\frac{\pi}{4}\sigma_{1}^{y}}e^{-i\frac{\pi}{2}S_{2}^{y}}e^{-i\frac{\pi}{4}\sigma_{3}^{y}}\left\vert \downarrow \downarrow \right\rangle \left\vert -3 \right\rangle.
\end{equation}

In the same way the probability observable $P_{s_{1},s_{3},m}$ for the state $\left\vert s_{1}s_{3} \right\rangle \left\vert m \right\rangle$ can be determined. By scanning the laser frequency, the auxiliary spin-$\frac{1}{2}$ ion comes into resonance and consequently scatter light. The auxiliary resonance frequency is defined depending of the spin states of the other three ions. The readout method cannot be applied directly for a general collective spin state with more than one spin-$3$ ion. For example, in a mixed spin system with two spin-$\frac{1}{2}$ ions and two spin-$3$ ions, the states $\left\vert \uparrow \uparrow \right\rangle \left\vert m,-m\right\rangle$ and $\left\vert \uparrow \uparrow \right\rangle \left\vert -m ,m\right\rangle$ create equal amounts of displacement to the equilibrium position of the auxiliary ion, such that both states are indistinguishable. For these more general cases, the state detection problem can be solved by (i) applying a spatially varying magnetic field gradient with $\partial^{2}B/\partial z^{2}\neq 0$ which allows one to distinguish the spin states due to the position dependence of the force, or (ii) a separating the entire ion crystal into smaller sub-crystal and transporting them into magnetic gradient detection zones \cite{WALTER2011,HUBER2010} which is a promising solution especially for segmented micro traps \cite{2DSS}. The very precise measurement of the position wave function of an ion was an experimental key element in experiments which could demonstrate the quantum random walk \cite{KIRCH2010}, was also used in other ion trap experiments with single ions and relies on the light ion interaction in either Raman \cite{POSCH2010} or bichromatic light fields \cite{KIRCH2010,DiracEq}. These methods can determine the wave packet position to an accuracy of a few percent of the ground state wavepacket extension $\Delta z_m$, but for an application here one would need to lock the laser light phases for position determination with respect to the initialization RF pulses on the ions, which has not been experimentally realized so far.

%%%%%%%%%%%%%%%%%%%%%%%%%%%%%%%%%%%%%%%%%%%%%%%%%%%%%%%%%%%%%%%%%%%%%%%
\section{Conclusion and Outlook}\label{Sec:conclusion}
%%%%%%%%%%%%%%%%%%%%%%%%%%%%%%%%%%%%%%%%%%%%%%%%%%%%%%%%%%%%%%%%%%%%%%%
We have proposed a method for the creation and manipulation of the spin-spin interactions in spin-$\frac{1}{2}$ ion crystal doped with high magnetic moment ions with spin $S=3$. It is shown that by tuning the frequency and direction of the oscillating magnetic field gradients various fundamental models in quantum magnetism of mixed spin systems can be realized. Because of the competing long-range spin-spin couplings the spin orders are extremely numerous even for spin system consisting of a small number of ions. We have proposed a technique for spin preparation and readout based on the frequency addressing of an indicator spin-$\frac{1}{2}$ ion in the presence of spatially varying magnetic field.

In future we will investigate the decoherence properties of such states, and how the quantum simulation is affected by the typical noise sources in an ion trap experiment. As the different quantum states are affected very differently by the expected dominating sources of noise, the ambient magnetic field fluctuations, we expect that states in decoherence-free subspaces, as for example $\left\vert\psi_{f,1}\right\rangle=(\left\vert\uparrow\uparrow\right\rangle\left\vert -1 \right\rangle + \left\vert\downarrow \downarrow \right\rangle \left\vert 1\right\rangle)/\sqrt{2}$, or $\left\vert\psi_{a,3}\right\rangle =(\left\vert\uparrow\downarrow\right\rangle +\left\vert\downarrow \uparrow \right\rangle)(\left\vert 3 \right\rangle + \left\vert -3 \right\rangle)/2$ with zero effective magnetic moment are preferred. A second future research theme is the combination of spin-dependent forces by magnetic gradients and with an advanced control of multiple segments in a micro trap \cite{SCHULZ2008,2DSS} which form the axial potential for the confinement of the ion crystal in the $z$-direction. As this potential may be shaped in non-trivial, non-harmonic way and even with multiple potential wells \cite{HAR2010,BROWN2011} the interspatial distances in the ion crystal become free parameters instead of being given by the Coulomb repulsion in an overall harmonic potential. Here, it was shown that $2\times N$ cluster states can be generated \cite{WUNDER2009} as a resource for one way quantum computing.

%%%%%%%%%%%%%%%%%%%%%%%%%%%%%%%%%%%%%%%%%%%%%%%%%%%%%%%%%%%%%%%%%%%%%%%%
\ack This work has been supported  the Bulgarian NSF grants VU-I-301/07 and D002-90/08.
%%%%%%%%%%%%%%%%%%%%%%%%%%%%%%%%%%%%%%%%%%%%%%%%%%%%%%%%%%%%%%%%%%%%%%%

\appendix

%%%%%%%%%%%%%%%%%%%%%%%%%%%%%%%%%%%%%%%%%%%%%%%%%%%%%%%%%%%%%%%%%%%%%%%
\section{Calculation of the eigenfrequencies and eigenvectors of the impurity doped ion crystal}\label{appendix}
%%%%%%%%%%%%%%%%%%%%%%%%%%%%%%%%%%%%%%%%%%%%%%%%%%%%%%%%%%%%%%%%%%%
In order to determine the spin-spin interactions we need to calculate the eigenfrequencies and eigenvectors of the impurity doped ion crystal.
The potential energy of collection of $N-K$ ions with mass $m$ and $K$ ions with mass $M$ is
\begin{eqnarray}
V&=&\frac{m\omega_{z}(m)^{2}}{2}\sum_{j=1}^{N-K}z_{j}^{2}+\frac{m\omega_{r}(m)^{2}}{2}\sum_{j=1}^{N-K}(x_{j}^{2}+y_{j}^{2})\nonumber \\ &&+\frac{M\omega_{z}(M)^{2}}{2}\sum_{k=1}^{K}z_{k}^{2}+\frac{M\omega_{r}(M)^{2}}{2}\sum_{k=1}^{K}(x_{k}^{2}+y_{k}^{2})\nonumber \\
&&+\frac{e^{2}}{8\pi\varepsilon_{0}}
\sum_{j,k=1\atop j\neq k}^{N}\frac{1}{\sqrt{(x_{j}-x_{k})^{2}+(y_{j}-y_{k})^{2}+(z_{j}-z_{k})^{2}}},\label{V}
\end{eqnarray}
where for simplicity of the notation we assume that the trapping confinements in the radial $x$-$y$ direction are equal, $\omega_{x}=\omega_{y}=\omega_{r}$. Here $\omega_{z}(m)\sim m^{-1/2}$ and $\omega_{z}(M)\sim M^{-1/2}$ are, respectively, the axial trap frequency for ion with mass $m$ and $M$. The radial oscillation frequency $\omega_{r}^{0}(a)\sim a^{-1}$ with $a=m,M$ is reduced by the axial trap frequency according to
\begin{equation}
\omega_{r}(a)=\omega_{r}^{0}(a)\left(1-\frac{\omega_{z}(a)^{2}}{2\omega_{r}^{0}(a)^{2}}\right)^{1/2},
\end{equation}
in lowest-order approximation \cite{RMP}. For small displacement the motional degrees of freedom in $x$, $y$, and $z$ directions are decoupled. The Hessian matrices $A_{ij}$ and $B_{ij}$ which describe the small oscillation of the ions around their equilibrium position in the axial $z$, and respectively, $x$-$y$ directions are given by
\begin{equation}
A_{ij}=\left\{\begin{array}{c}
1+2\sum_{j=1 \atop j\neq p}^{N}\frac{1}{\left\vert u_{j}-u_{p}\right\vert ^{3}},\ (i=j), \\
\frac{-2}{|u_{i}-u_{j}|^{3}},\ (i\neq
j),
\end{array}\right.
\end{equation}
\begin{equation}
B_{ij}=\left\{\begin{array}{c}
1-\frac{\alpha^{2}}{2}-\alpha ^{2} \sum_{j=1\atop j\neq p}^{N}\frac{1}{\left\vert u_{j}-u_{p}\right\vert ^{3}},\ (i=j\neq j_{M}), \\
\frac{1}{\mu}-\frac{\alpha^{2}}{2}-\alpha ^{2}\sum_{j=1\atop j\neq p}^{N}\frac{1}{\left\vert u_{j}-u_{p}\right\vert ^{3}},\ (i=j=j_{M}),\\
\frac{\alpha ^{2}}{|u_{i}-u_{j}|^{3}},\ (i\neq j).
\end{array}\right.
\end{equation}
Here we have introduced the mass ratio $\mu=M/m$ and anisotropy parameter $\alpha=\omega_{z}(m)/\omega_{r}^{0}(m)$ and $u_{j}$ being the dimensionless equilibrium position of the $j^{th}$ ion. In the above expressions the index $j_{M}$ denotes the position of the $j^{th}$ ion with mass $M$. The eigenfrequencies $\omega_{z,n}=\omega_{z}(m)\sqrt{\lambda_{n}}$ and eigenvectors $\textbf{b}_{n}^{z}$ ($n=1,2,\ldots,N$) in the $z$ direction are given by the diagonalization of the following matrix
\begin{equation}
\tilde{A}_{ij}=\left\{\begin{array}{c}
A_{ij},\ (i=j\neq j_{M}), \\
A_{ij}/\mu,\ (i=
j=j_{M}),\\
A_{ij},\ (i\neq j\neq j_{M}),\\
A_{ij}/\sqrt{\mu},\ (i\, \textrm{or}\, j=j_{M},\ i\neq j),\\
A_{ij}/\mu,\ (i_{M}\neq j_{M}),
\end{array}\right.
\end{equation}
such that $\sum_{j=1}^{N}\tilde{A}_{ij}b_{j,n}^{z}=\lambda_{n}b_{i,n}^{z}$. In the same way the eigenfrequencies $\omega_{q,n}=\omega_{r}^{0}(m)\sqrt{\gamma_{n}}$ and eigenvectors $\textbf{b}_{n}^{q}$ ($q=x,y$) in the radial direction are obtained by the diagonalization of the following matrix
\begin{equation}
\tilde{B}_{ij}=\left\{\begin{array}{c}
B_{ij},\ (i=j\neq j_{M}), \\
B_{ij}/\mu,\ (i=
j=j_{M}),\\
B_{ij},\ (i\neq j\neq j_{M}),\\
B_{ij}/\sqrt{\mu},\ (i\, \textrm{or}\, j=j_{M},\ i\neq j),\\
B_{ij}/\mu,\ (i_{M}\neq j_{M}),
\end{array}\right.
\end{equation}
with $\sum_{j=1}^{N}\tilde{B}_{ij}b_{j,n}^{q}=\gamma_{n}b_{i,n}^{q}$

\end{document}